\documentclass[12pt]{article}

\ifx\pdfoutput\undefined
\usepackage[dvips,bookmarks]{hyperref}
\else
\usepackage{hyperref}
\fi
\hypersetup{colorlinks=false,bookmarksopen,bookmarksnumbered,citecolor=blue,
   pdfstartview=FitH}

\usepackage{latexsym}
\usepackage{amssymb,amsfonts,amsmath}
\usepackage{graphicx} 
\usepackage{indentfirst}
\usepackage{bbm}
\usepackage{amssymb}
\usepackage{verbatim}
\usepackage{amsmath, amsthm,amssymb}
\usepackage{mathrsfs}
\usepackage{hyperref}
\usepackage{amsfonts}
\usepackage{dsfont}
\usepackage{slashed, tensor}
\usepackage{booktabs}
\usepackage{graphicx}
\usepackage{mathrsfs}

\parskip=\medskipamount

\newcommand {\cA}{{\cal A}}
\newcommand {\cB}{{\cal B}}
\newcommand {\cC}{{\cal C}}
\newcommand {\cD}{{\cal D}}

\newcommand {\cF}{{\cal F}}

\newcommand {\cM}{{\cal M}}
\newcommand {\cN}{{\cal N}}

\newcommand {\cS}{{\cal S}}

\newcommand {\cW}{{\cal W}}

\def\a{\alpha}

\def\b{\beta}
\def\c{\chi}
\def\d{\delta}

\def\f{\phi}
\def\g{\gamma}

\def\k{\kappa}
\def\l{\lambda}

\def\q{\theta}

\def\s{\sigma}
\def\t{\tau}

\def\F{\Phi}

\def\O{\Omega}

\def\tr{{\rm tr}}

\def\ri{{\rm i}}
\def\re{{\rm e}}

\newcommand{\ve}{\varepsilon}                            

\newcommand{\pa}{\partial}                           
\newcommand{\hf}{\frac12}

\newcommand{\be}{\begin{equation}}
\newcommand{\ee}{\end{equation}}
\newcommand{\bea}{\begin{eqnarray}}
\newcommand{\eea}{\end{eqnarray}}
\newcommand{\non}{\nonumber}
\newcommand{\ba}{\begin{array}}
\newcommand{\ea}{\end{array}}

\newcommand{\bm}[1]{\mbox{\boldmath$#1$}}

\def\double #1{#1{\hbox{\kern-2pt $#1$}}}

\newcommand{\sSp}{\mathsf{Sp}}
\newcommand{\sSU}{\mathsf{SU}}

\newcommand{\sSO}{\mathsf{SO}}
\newcommand{\sO}{\mathsf{O}}
\newcommand{\sU}{\mathsf{U}}

\newcommand{\sOSp}{\mathsf{OSp}}

\newcommand{\bsubeq}{\begin{subequations}}
\newcommand{\esubeq}{\end{subequations}}

\numberwithin{equation}{section}

\begin{document}

\begin{titlepage}
\begin{flushright}
July, 2015
\end{flushright}
\vspace{2mm}

\begin{center}
{\Large \bf On superconformal Chern-Simons-matter  theories
\\[4pt]
 in $\bm{\cN=4}$ superspace}
\end{center}

\begin{center}

{\bf
Sergei M. Kuzenko and Igor B.
Samsonov\footnote{On leave from Tomsk Polytechnic University, 634050
Tomsk, Russia.}
}

{\footnotesize{
{\it School of Physics M013, The University of Western Australia\\
35 Stirling Highway, Crawley W.A. 6009, Australia}} ~\\
}

\end{center}

\begin{abstract}
\baselineskip=14pt
In three dimensions, every known $\cN=4$ supermultiplet has an off-shell completion. 
However, there is no off-shell $\cN=4$ formulation for the known extended superconformal Chern-Simons  (CS) theories with eight and more supercharges. 
To achieve a better understanding of this issue, we provide $\cN=4$
superfield realisations for the equations of motion which correspond 
to various $\cN=4$  and $\cN=6$  superconformal CS theories, including 
the Gaiotto-Witten theory and the ABJM theory. 
These  superfield realisations demonstrate that 
the  superconformal CS theories with $\cN\geq 4$
(except for the Gaiotto-Witten theory)
require a reducible long $\cN=4$ vector multiplet, 
from which the standard left and right $\cN=4$ vector multiplets
are obtained by constraining the field strength to be either self-dual or anti self-dual. 
Such a long multiplet naturally originates upon reduction of 
any off-shell $\cN>4$ vector multiplet to $\cN=4$ superspace. 
For the long $\cN=4$ vector multiplet we develop a prepotential formulation. 
It makes use of two prepotentials being subject to the constraint 
which defines the so-called hybrid projective multiplets
introduced in the framework of $\cN=4$ supergravity-matter systems
in arXiv:1101.4013.
We also couple $\cN=4$ superconformal CS theories to $\cN=4$ conformal supergravity. 
\end{abstract}

\end{titlepage}

\newpage
\renewcommand{\thefootnote}{\arabic{footnote}}
\setcounter{footnote}{0}

\newpage

\section{Introduction}

Since the 2004 work by Schwarz  \cite{Schwarz}, much progress
has been achieved in the construction of extended 
superconformal Chern-Simons-matter (CS) theories in three dimensions (3D).
The famous CS theories with $\cN=8$ \cite{BLG1,BLG2,Gustavsson},
$\cN=6$ \cite{ABJM,HLLLP2,ABJ} and $\cN=4$  \cite{GW,HLLLP} supersymmetry 
have been proposed. 

All superconformal CS theories with $\cN>3$ can be realised as 
special off-shell $\widehat \cN$-extended Chern-Simons-matter  systems, 
where $\widehat \cN \leq 3$. In such realisations in terms of 
$\widehat \cN$-extended superfields, $\cN- \widehat \cN$ supersymmetries are hidden. Of course, the $\widehat \cN=3$ realisation \cite{BILPSZ} is the most 
powerful, since it allows one to keep manifest the maximal amount of supersymmetry. 
The special feature of the three cases $\cN=1$, $\cN=2$ and $\cN=3$ is that 
the off-shell supersymmetric {\it pure} CS action exists for any gauge group. 
However, no  $\cN \geq 4$
supersymmetric CS action can be constructed (for a
recent proof, see \cite{KN14}), although abelian $\cN=4$ BF
couplings are abundant \cite{BrooksG}.
In this regard, especially paradoxical is the situation with $\cN=4$ supersymmetry. 
Every 3D $\cN=4$ supermultiplet admits an off-shell realisation. 
There exist off-shell formulations for various
3D $\cN=4$ supersymmetric theories, including 
the Yang-Mills theories with Poincar\'e \cite{Zupnik98,Zupnik99}
and anti-de Sitter supersymmetry \cite{KT-M14},  
the most general  $\s$-models with Poincar\'e \cite{KPT-MvU}
and anti-de Sitter supersymmetry \cite{BKT-M}, 
and general supergravity-matter systems \cite{KLT-M11}.
However, it is impossible
to construct a $\cN=4$ supersymmetric CS action, at least 
in terms of the standard vector multiplets and hypermultiplets.

Since there is no way to realise the known $\cN \geq 4$ superconformal CS theories
in terms of $\cN=4$ superfields off the mass shell, in this note we would like to analyse 
a simpler problem. We will only formulate the equations of motion for 
 $\cN \geq 4$ superconformal CS theories in $\cN=4$ superspace. 
Similar on-shell realisations in $\cN=6$ and  $\cN=8$ superspaces have been given 
in \cite{SW8,SW6,GGHN}.

This note is organised as follows. In section 2 we consider general 
$\cN=4$ superconformal CS theories and show that the hypermultiplet equations of motion require consistency conditions which impose non-trivial constraints on the gauge group and its representation to which the hypermultiplet belongs.  
In sections 3 and 4 we present the $\cN=4$ superfield equations of motion for the Gaiotto-Witten and ABJM theories, respectively. For these models we  also construct their supercurrents and other conserved current multiplets. 
Section 5 is devoted to the prepotential formulation 
for the large $\cN=4$ vector multiplet. 
In conclusion, we discuss the structure of the long $\cN=4$ vector multiplet from  $\cN=3$ superspace perspective. In appendix A we give a proof that the constraints on the gauge group derived in section 2 are equivalent to the fundamental identity of Gaiotto and Witten \cite{GW}. In appendix B we review the structure of the 
$\cN$-extended vector multiplet coupled to conformal supergravity.


\section{$\cN=4$ superconformal CS theories}
\label{SectN4}

The $\cN=4$ Minkowski superspace can be parametrised by  coordinates
$z^{\cal A} = (x^{\a\b}, \theta_{i \tilde i}^\alpha)$. Here 
$x^{\a\b}= x^{(\a\b)}$ are the bosonic coordinates, where $\a, \b =1,2$ 
are spinor indices.\footnote{The variables  $x^{\a\b}$ are related to the 
coordinates $x^m$ of Minkowski space ${\mathbb M}^{3}$
by the rule $x^{\a\b} = x^m (\g_m)^{\a\b}$, with $(\g)^{\a\b}$ the gamma-matrices
with upper spinor indices
The partial derivatives $\pa_{\a\b}$ are defined similarly, 
$\pa_{\a\b} =  (\g^m)_{\a\b}\pa_m$, such that 
$\pa_{\a \b} x^{\g\d} = -2 \d_\a{}^{(\g} \d_\b{}^{\d)} $.
Our two-component spinor notation and conventions, including the 
definition of the gamma-matrices, follow \cite{KPT-MvU,KLT-M11}. } 
The Grassmann coordinates 
$\theta_{i\tilde i}^\alpha$ carry 
two isospinor indices,  $i=1,2$ and $\tilde i=1,2$,  
which correspond to the subgroups $\sSU(2)_{\rm L}$ and $\sSU(2)_{\rm R}$ 
of the $\cN=4$  $R$-symmetry group  $\sSU(2)_{\rm L} \times \sSU(2)_{\rm R}$. 
The spinor covariant derivatives $ D^{i\tilde i}_\alpha$ satisfy the 
anti-commutation relations
$
\{D_\a^{i{\tilde i}},D_\b^{j{\tilde j}}\}=2\ri\ve^{ij}\ve^{{\tilde i}{\tilde j}}\pa_{\a\b}
$. 

To describe a non-abelian $\cN=4$ vector multiplet, 
we introduce gauge covariant derivatives 
$\cD_{\cA}=(\cD_{\alpha\b}, \cD^{i\tilde i}_\alpha)
= D_\cA +\ri V_\cA$, where 
$D_{\cA}=(\partial_{\alpha\b}, D^{i\tilde i}_\alpha)$ denotes
the covariant derivatives of $\cN=4$ superspace.  
The gauge connection $V_\cA$ takes its values in the Lie algebra of the gauge 
group $G$.
Given a matter multiplet $\F$ belonging to some representation of the 
gauge group, the gauge transformation laws of $V_\cA$ and $\F$
are as follows:
\begin{subequations} \label{3.1}
\bea
{\cal D}'_\cA &=& \re^{\ri \tau} {\cal D}_\cA \re^{-\ri\tau }~, \label{3.1a}\\
\F'&=& \re^{\ri\tau} \F~. \label{3.1b}
\eea
\end{subequations}
Here the Lie-algebra-valued gauge parameter $\tau (z) $ is Hermitian, 
$\t^\dagger =\t$, but otherwise unconstrained. 

In 3D $\cN$-extended supersymmetry, one can impose a universally looking constraint
to describe a vector multiplet, see Appendix \ref{AppendB}.
In the $\cN=4$ case the constraint amounts to
\be
\{{\cal D}_\a^{i\tilde i},{\cal D}_\b^{j\tilde j}\}
=2\ri\ve^{ij}\ve^{\tilde i\tilde j}{\cal D}_{\a\b} + \varepsilon_{\alpha\beta} \varepsilon^{ij}
 \cW^{\tilde i \tilde j} 
+\varepsilon_{\alpha\beta} \varepsilon^{\tilde i \tilde j } \cW^{ij}
~.
\label{N4-algebra}
\ee
The field strengths $\cW^{ij} = \cW^{(ij)}$ and 
$\cW^{\tilde i \tilde j} = \cW^{(\tilde i \tilde j )}$ 
are Hermitian in the sense that 
$(\cW^{ij})^\dagger = \cW_{ij} = \ve_{ik} \ve_{jl} \cW^{kl}$, 
and similarly for $\cW^{\tilde i \tilde j} $.
They are subject to the  Bianchi identities
\begin{subequations}
\label{W-constraints}
\bea
{\cal D}_\alpha^{\tilde i( i }\cW^{jk)} &=&0~,\\
{\cal D}_\alpha^{i(\tilde i} \cW^{\tilde j \tilde k)}&=&0~.
\eea
\end{subequations}

The vector multiplet described by the constraint \eqref{N4-algebra}
is reducible and, therefore, will be called ``long.'' 
There exist two irreducible off-shell $\cN=4$ vector multiplets
 \cite{BrooksG,Zupnik98,Zupnik99}, which are obtained from \eqref{N4-algebra}
by imposing additional constraints.  
 Following the terminology of \cite{KLT-M11}, 
 the left vector multiplet is subject to the additional constraint%
\begin{subequations}
 \bea
 \cW^{\tilde i \tilde j} =0~. \label{3.4a}
 \eea
The right vector multiplet is obtained by setting 
\bea
 \cW^{i  j} =0~. \label{3.4b}
 \eea
\end{subequations}
In what follows, we will work with the long vector multiplet \eqref{N4-algebra}
due to the following two reasons: (i) as will be shown below, 
it naturally corresponds to the $\cN \geq 4$
superconformal CS theories; and (ii) it is obtained by reducing the off-shell 
$\cN >4$ vector multiplets to $\cN=4$ superspace.

Similar to the left and right vector multiplets, there are two inequivalent
$\cN=4$  hypermultiplets, left and right ones. 
In this paper, we will be interested in {\it on-shell} hypermultiplets.
The left hypermultiplet is described by a left  isospinor $q^i=(q^i_a)$
(which is viewed in this section as a column vector) 
and its conjugate $\bar q_i= (q^i )^\dagger$.
The right hypermultiplet is described by a right  isospinor  
$q^{\tilde i}=(q^{\tilde i}_{\tilde a})$ and its conjugate
$\bar q_{\tilde i}= ( q^{\tilde i} )^\dagger$.
Both the left and right hypermultiplets are assumed to interact with the long vector multiplet.  In general they belong to  different representations of the gauge group $G$,
with the generators $(T_A)_a{}^b$ and $(\tilde T_{A})_{\tilde a}{}^{\tilde b}$,
respectively. The hypermultiplet equations of motion have the form
\begin{subequations}
\label{9}
\bea
\label{9a}
{\cal D}_\alpha^{\tilde i (i}q^{j)} =0~, & \qquad  & 
{\cal D}_\alpha^{\tilde i (i}\bar q^{j)}=0~,\\
{\cal D}_\alpha^{i(\tilde i }q^{\tilde j)} = 0 ~, & \qquad    &
{\cal D}_\alpha^{i (\tilde i }\bar q^{\tilde j)}=0~,
\label{9b}
\eea
\end{subequations}
and are similar to the constraints introduced by Sohnius \cite{Sohnius} 
to describe the $\cN=2$ hypermultiplet in four dimensions. 
The crucial difference of these equations from their $\cN=3$ counterparts is that they require the following consistency conditions
\be
\cW^{(ij}q^{k)} =0 ~,\qquad
\cW^{(\tilde i \tilde j} q^{\tilde k)}=0~.
\label{int-cond}
\ee
In the case of $\cN=3$ superconformal CS theories with matter, no 
restriction on the gauge group and its representation occur, 
see \cite{BILPSZ} for more details. 

Up to now, our consideration was completely general. 
In what follows we restrict ourselves to superconformal theories. 
In this case the gauge multiplet cannot have independent degrees of freedom 
as the Yang-Mills coupling in not permitted. On the equations of motion the field
 strengths should be expressed in terms of hypermultiplets. 
A large family  of superconformal theories is described by the equations of motion for the field strengths which are bilinear in hypermultiplets
\begin{subequations}\label{WW}
\bea
\cW^{ij}_A &=& \ri \kappa\,g_{AB}\, \bar q^{(i} T^B  q^{j)}~,\label{W}\\
\cW^{\tilde i \tilde j}_A &=& \ri \tilde \kappa\,g_{AB}\, \bar q^{(\tilde i} \tilde T^B q^{\tilde j)}~,
\label{tilde-W}
\eea
\end{subequations}
where $\kappa$ and $\tilde \kappa$ are some dimensionless coefficients and $g_{AB}$
is an invariant quadratic form on the Lie algebra of the gauge group $G$. 
The consistency conditions (\ref{int-cond}) lead to the following equations
\begin{subequations}
\bea
\bar q^{a(i}q^j_b q^{k)}_c\, g_{AB}\,(T^A)_a{}^b (T^B)_d{}^c &=&0~,
\label{constraint}\\
\bar q^{\tilde a(\tilde i}q^{\tilde j}_{\tilde b} q^{\tilde k)}_{\tilde c}\, 
g_{AB}\,(\tilde T^{A})_{\tilde a}{}^{\tilde b} (\tilde T^{B})_{\tilde d}{}^{\tilde c} &=&0~.
\eea
\end{subequations}
These equations require the generators to obey the relations
\begin{subequations}
\label{15}
\bea
g_{AB}\,(T^A)_a{}^{(b} (T^B)_d{}^{c)}&=&0~,\label{15a}\\
g_{AB}\,(\tilde T^{A})_{\tilde a}{}^{(\tilde b} (\tilde T^{B})_{\tilde d}{}^{\tilde c)} &=&0~,
\eea
\end{subequations}
which are strong constraints on the possible gauge group $G$ and 
its representations. These relations are, in fact, equivalent to the fundamental identity for the generators of the gauge group derived in \cite{GW}
(see Appendix \ref{AppA} for the proof).

The dynamical system under consideration is characterised by the supercurrent 
(compare with \cite{BKS2}) 
\be
J= \bar q_i  q^i -  \bar q_{\tilde i} q{}^{\tilde i} ~,
\ee
which obeys the conservation equation \cite{KNT-M,KN14,BKS2}
\be
D^{\alpha\, (\tilde i (i } D_\alpha^{j)\tilde j )}
J =0~,  
\label{21.a}
\ee
as a consequence of the equations of motion \eqref{9} and \eqref{WW}.


\section{The equations of motion for the Gaiotto-Witten theory} \label{GWsection}

In the previous section we have provided the $\cN=4$ superfield description for the general 
$\cN=4$ superconformal CS theories studied in \cite{HLLLP}.
The Gaiotto-Witten theory \cite{GW} is a special member of this family.
This theory has only one type of hypermultiplets, $q^i$, 
and no right hypermultiplets, 
$q^{\tilde i}=0$. Then we should also have ${\cal W}^{\tilde i \tilde j}=0$, 
as a consequence of (\ref{tilde-W}), and the vector multiplet becomes short, the left one.  
The remaining superfields $q^i$ and ${\cal W}^{ij}$ obey the equations of motion (\ref{9a}) and (\ref{W}). Now we will show that the constraint 
\eqref{15a}
is satisfied for the Gaiotto-Witten theory \cite{GW}. 

This theory has two field strengths  ${\cal W}_{\frak L}^{ij}$ and ${\cal W}_{\frak R}^{ij}$ 
associated with a gauge group of the form $G=G_{\frak L}\times G_{\frak R}$ 
that possesses a representation compatible with eq. \eqref{15a}
(see also the discussion in Appendix \ref{AppA}). One admissible choice is
$G= \sU (M) \times \sU(N)$, and the hypermultiplet transforms in the bi-fundamental representation of $G$. 
Only this case is considered in the present section. 
The gauge transformation laws of these superfields are 
\begin{subequations}
\bea
 q'{}^i &=& \re^{\ri\tau_{\frak L}} q^i \re^{ - \ri \tau_{\frak R}}~,\qquad \qquad
 \bar q'{}^i = \re^{\ri\tau_{\frak R}} \bar q^i \re^{ - \ri  \tau_{\frak L}}~, \label{14.a}\\
 {\cal W}'^{ij}_{\frak L} &=&  \re^{\ri \tau_{\frak L}}{\cal W}_{\frak L}^{ij} 
 \re^{-\ri \tau_{\frak L}}~,\qquad
{\cal W}'^{ij}_{\frak R} =  \re^{\ri \tau_{\frak R}}{\cal W}_{\frak R}^{ij} \re^{-\ri \tau_{\frak R}}
\label{gauge-tr}
\eea
\end{subequations}
where the gauge parameters $\tau_{\frak L} (z) $ and $\tau_{\frak R}(z)$ are 
Hermitian matrices taking their values in the Lie algebras of the gauge groups 
$G_{\frak L}$ and $G_{\frak R}$, respectively. 

The equations of motion for the hypermultiplets (\ref{9a}) 
and the vector multiplet (\ref{W}) become
\begin{subequations}
\label{GW-eom}
\bea
{\cal D}_\alpha^{\tilde i (i}q^{j)} &=& 0~,\qquad \quad ~~
{\cal D}_\alpha^{\tilde i (i}\bar q^{j)}=0~,\label{q-eom}\\
{\cal W}_{\frak L}^{ij} &=& \ri \kappa \, q^{(i} \bar q^{j)}~,\qquad
{\cal W}_{\frak R}^{ij} = \ri \kappa \, \bar q^{(i} q^{j)}~.\label{W-eom}
\eea
\end{subequations}
Here the covariant derivatives act on the hypermultiplet superfields by the rule
\be
{\cal D}_\alpha^{\tilde i i}q^{j} =  D_\alpha^{\tilde i i}q^{j}
+\ri V_{\frak L\,\alpha}^{\tilde i i } q^j -\ri q^j V_{\frak R\,\alpha}^{\tilde i i}~, 
\qquad
{\cal D}_\alpha^{\tilde i i}\bar q^{j} =  D_\alpha^{\tilde i i}\bar q^{j}
+\ri V_{\frak R\,\alpha}^{\tilde i i }\bar q^j -\ri \bar q^j V_{\frak L\,\alpha}^{\tilde i i}~,
\ee
in accordance with the transformation laws \eqref{14.a}.
For the equations of motion (\ref{q-eom}), 
the consistency conditions (\ref{int-cond}) take the form 
\be
{\cal W}_{\frak L}^{(ij} q^{k)} - q^{(k}{\cal W}_{\frak R}^{ij)} =0~,\qquad
{\cal W}_{\frak R}^{(ij} \bar q^{k)} - \bar q^{(k}{\cal W}_{\frak L}^{ij)} =0
\ee
and are identically satisfied for the superfield strengths  (\ref{W-eom}).

In concluding this section we construct the $\cN=4$ supercurrent $J$ and $\sU(1)$ flavour current multiplet $L^{ij}$ in the Gaiotto-Witten theory
\be
J = \tr\, (q^i \bar q_i) ~,\qquad
L^{ij} = \ri\, \tr\,  (q^{(i}  \bar q^{j)})~,
\label{J-L}
\ee
which obey the conservation equations (see \cite{BKS2} for more details)
\bea
D^{\alpha\, (\tilde i (i } D_\alpha^{j)\tilde j )}
J =0~,  \qquad
D^{\tilde i (i}L^{jk)}&=&0~,
\eea
as a consequence of the equations of motion (\ref{GW-eom}).
Of course, the flavour current multiplet is non-trivial only for the gauge group which possesses the $\sU(1)$ factor.
The two- and three-point correlation functions of the 
supercurrent and flavour current multiplets in general $\cN=4$ 
superconformal field theories 
were studied in \cite{BKS2}.

\section{The equations of motion for the ABJM theory}

Before presenting our $\cN=4$ superfield realisation for the $\cN=6$ superconformal
CS theory proposed in \cite{ABJM} and known as the ABJM theory, 
we would like to make some preliminary comments. In three dimensions, 
the $\cN$-extended vector multiplet can be formulated in $\cN$-extended superspace
and is off-shell  \cite{GGHN} (see also \cite{BKNT-M1}). A brief review of 
the $\cN$-extended vector multiplet coupled to conformal supergravity
is given in Appendix \ref{AppendB}. In the flat case, 
every $\cN>4$ vector multiplet can be reduced to $\cN=4$ Minkowski superspace, 
resulting in the long $\cN=4$ vector multiplet 
coupled to several additional constrained superfields. 
In particular, it can be shown that 
the $\cN=6 \to \cN=4$  reduction for the field strength $\cW^{IJ} = -\cW^{JI}$
of the $\cN=6$ vector multiplet\footnote{Using the isomorphism 
$\sSU(4) \cong \sSO(6)/{\mathbb Z}_2$, the $\cN=6$ vector multiplet can be described 
in the $\sSU(4)$ notation \cite{SW6}.} 
(with $I,J$ being $\sSO(6)$ indices, see Appendix \ref{AppendB})
leads to the following  
$\cN=4$ superfields
\be
{\cal W}^{ij}~,\quad
{\cal W}^{\tilde i \tilde j}~,\quad
{\cal Y}^{i\tilde i }~,\quad
{\cal Z}^{i\tilde i}~, \quad
{\cal S}~.
\label{N6-superfields}
\ee
It may be shown that the $\cN=6$  Bianchi identity \eqref{VMBI}
is equivalent to the following constraints on the above $\cN=4$ superfields:
\begin{subequations}
\bea
&&{\cal D}_\a^{\tilde i(i}{\cal W}^{jk)} =0~,\qquad ~
 {\cal D}_\a^{i(\tilde i}{\cal W}^{\tilde j \tilde k) }=0~,\\
&&{\cal D}_\a^{(\tilde i( i }{\cal Y}^{j)\tilde j)}  
 =0~, \qquad {\cal D}_\a^{(\tilde i( i }{\cal Z}^{j)\tilde j)} =0~,
 \label{30}\\
&&{\cal D}^{\alpha(\tilde i (i} {\cal D}^{j)\tilde j)}_\alpha
{\cal S}=[{\cal Y}^{(\tilde i(i},{\cal Z}^{j)\tilde j)}]
~.\label{31}
\eea
\end{subequations}
The constraints \eqref{30} tell us that ${\cal Y}^{i\tilde i }$ and $ {\cal Z}^{i\tilde i}$
are examples of the so-called hybrid supermultiplets introduced
for the first time  in \cite{KLT-M11} in the framework of 
general $\cN=4$ supergravity-matter systems. Eq. \eqref{31} may be interpreted 
as the condition that $\cS$ is a  hybrid linear superfield. 

Unlike the theory studied in section \ref{GWsection},
now we consider the case when both hypermultiplets $q^i$ and $q^{\tilde i}$ have non-trivial dynamics. Here we assume that the hypermultiplets  transform in 
the bi-fundamental representation of the gauge group 
$G=G_{\frak L} \times G_{\frak R} = \sU(M) \times \sU(N)$, 
\begin{subequations}
\bea
q'^i = e^{\ri\tau_{\frak L}} q^i e^{ - \ri \tau_{\frak R}}~,\qquad
\bar q'^i = e^{\ri\tau_{\frak R}} \bar q^i e^{ - \ri  \tau_{\frak L}}~,\\
q'^{\tilde i} = e^{\ri\tau_{\frak L}} q^{\tilde i} e^{ - \ri \tau_{\frak R}}~,\qquad
\bar q'^{\tilde i} = e^{ \ri\tau_{\frak R}} \bar q^{\tilde i} e^{ - \ri \tau_{\frak L}}~,
\eea
\end{subequations}
where the gauge superfield parameters $\t_{\frak L} (z) $ and $\t_{\frak R} (z) $
are Hermitian and otherwise unconstrained.

Due to the structure of the gauge group, $G=G_{\frak L} \times G_{\frak R}$, 
there are two long $\cN=4$ vector multiplets and the corresponding field strengths. 
We have the field
 strengths ${\cal W}^{ij}_{\frak L}$ and ${\cal W}^{\tilde i\tilde j}_{\frak L}$ which take values in the Lie algebra of the gauge group $G_{\frak L}$ and similar ones, ${\cal W}^{ij}_{\frak R}$ and ${\cal W}^{\tilde i\tilde j}_{\frak R}$, which correspond to the gauge group $G_{\frak R}$. They transform in the adjoint representations of these groups
\begin{subequations}
\bea
{\cal W}'^{ij}_{\frak L} = e^{ \ri \tau_{\frak L}}{\cal W}_{\frak L}^{ij} e^{ -\ri \tau_{\frak L}}~,\qquad
{\cal W}'^{\tilde i\tilde j}_{\frak L} = e^{ \ri \tau_{\frak L}}{\cal W}_{\frak L}^{\tilde i\tilde j}e^{ -\ri \tau_{\frak L}}~,\\
{\cal W}'^{ij}_{\frak R} = e^{ \ri \tau_{\frak R}}{\cal W}_{\frak R}^{ij}e^{ -\ri \tau_{\frak R}}~,\qquad
{\cal W}'^{\tilde i\tilde j}_{\frak R} =  e^{\ri \tau_{\frak R}}{\cal W}_{\frak R}^{\tilde i\tilde j}e^{-\ri \tau_{\frak R}}~.
\eea
\end{subequations}

The natural generalisation of the equations of motion (\ref{GW-eom}) reads
\begin{subequations}
\label{ABJM-eom}
\bea
&&{\cal D}_\alpha^{\tilde i (i}q^{j)} = 0~,\qquad
{\cal D}_\alpha^{i (\tilde i} q^{\tilde j)}=0~,\label{qq-eom}\\
&&{\cal W}_{\frak L}^{ij} = \ri \kappa \, q^{(i} \bar q^{j)}~,\quad
{\cal W}_{\frak R}^{ij} = \ri \kappa \, \bar q^{(i} q^{j)}~,\quad
{\cal W}_{\frak L}^{\tilde i \tilde j} = \ri \kappa \, q^{(\tilde i} \bar q^{\tilde j)}~, 
\quad
{\cal W}_{\frak R}^{\tilde i\tilde j} = \ri \kappa \, \bar q^{(\tilde i} q^{\tilde j)}~.~~~~~
\label{WW-eom}
\eea
\end{subequations}
The consistency conditions \eqref{15}
are automatically satisfied  for these  equations,
\be
{\cal W}_{\frak L}^{(ij} q^{k)} - q^{(k}{\cal W}_{\frak R}^{ij)} =0~,\qquad
{\cal W}_{\frak L}^{(\tilde i\tilde j} q^{\tilde k)} - q^{(\tilde k}{\cal W}_{\frak R}^{\tilde i\tilde j)} =0~.
\ee

The ABJM theory is $\cN=6$ superconformal. Therefore, there should 
 exist hypermultiplet composites that realise 
the superfields  ${\cal Y}^{i\tilde i }$, ${\cal Z}^{i\tilde i} $ and ${\cal S}$
in (\ref{N6-superfields}) on the mass shell. 
Since we have two gauge groups,  $G_{\frak L}$ and  $G_{\frak R}$, 
the number of superfields (\ref{N6-superfields}) is doubled. 
We will distinguish them by attaching the subscripts $\frak L$ and $\frak R$ to them. 
It is clear that they should be expressed in terms of the hypermultiplet superfields. 
Indeed, the expressions for ${\cal W}^{ij}$ and ${\cal W}^{\tilde i \tilde j}$ 
are given by (\ref{WW-eom}). For the remaining superfields we find
\begin{subequations} \label{5.7}
\bea
&&{\cal Y}_{\frak L}^{i\tilde i} = 2\ri \kappa (q^i \bar q^{\tilde i} + q^{\tilde i} \bar q^i)~,\qquad
{\cal Y}_{\frak R}^{i\tilde i} = 2\ri \kappa (\bar q^i  q^{\tilde i} + \bar q^{\tilde i} q^i)~,\\ 
&&{\cal Z}_{\frak L}^{i\tilde i} =2 \kappa ( q^i \bar q^{\tilde i} - q^{\tilde i}\bar q^i)~,\qquad
{\cal Z}_{\frak R}^{i\tilde i} =2\kappa( \bar q^i  q^{\tilde i} - \bar q^{\tilde i} q^i)~,\\
&&
{\cal S}_{\frak L} = \kappa( q^i\bar q_i - q^{\tilde i}\bar q_{\tilde i})~,\qquad \quad
{\cal S}_{\frak R} = \kappa(\bar q_i q^i - \bar q_{\tilde i} q^{\tilde i})~.
\eea
\end{subequations}
These superfields do satisfy the $\cN=6$ Bianchi identities
 (\ref{30}) and (\ref{31}) on the hypermultiplet equations of motion (\ref{qq-eom}).

Since the ABJM theory is $\cN=6$ superconformal, it should possess
a number of conserved currents which form the $\cN=6 $ supercurrent multiplet $J^{IJ} = -J^{JI}$
\cite{BKNT-M1,KNT-M,BKS1}. Upon reduction to $\cN=4$ superspace, 
the $\cN=6 $ supercurrent may be shown to lead to the following constrained $\cN=4$ multiplets:
(i) the $\cN=4$ supercurrent $J$; (ii) two $\sU(1)$ flavour current multiplets
$L^{ij}  $ and $L^{\tilde i\tilde j} $; two $\sSO(4)$ vectors 
$A^{i\tilde i} $ and $B^{i\tilde i}$. 
In the  the ABJM theory, these multiplets should be given as hypermultiplet composites.
Their explicit form is as follows:
\begin{itemize}
\item the $\cN=4$ supercurrent
\be
J=\tr( q^i \bar q_i)  - \tr( q^{\tilde i} \bar q_{\tilde i})
\label{J-ABJM}
\ee
obeying the conservation equation \eqref{21.a};
\item the  $\sU(1)$ flavour current multiplets
\be
L^{ij} = \ri\, \tr( q^{(i}  \bar q^{j)}) ~,\qquad
L^{\tilde i\tilde j} = \ri\, \tr(  q^{(\tilde i}  \bar q^{\tilde j)})
\label{L-ABJM}
\ee
obeying  the conservation laws
\be
D_\a^{\tilde i (i}L^{jk)}=0~,\qquad
D_\a^{i (\tilde i}L^{\tilde j\tilde k)}=0~;
\ee
\item the $\sSO(4)$ vectors 
\be
A^{i\tilde i} = \ri\, \tr(q^i \bar q^{\tilde i}) + \ri\, \tr(\bar q^i q^{\tilde i})~,
\qquad
B^{i\tilde i}=\tr(q^i \bar q^{\tilde i}) - \tr(\bar q^i q^{\tilde i})~,
\label{A-ABJM}
\ee
which obey the same conservation equations
\be
D_\a^{(\tilde i (i} A^{j)\tilde j)}=0~, \qquad D_\a^{(\tilde i (i} B^{j)\tilde j)}=0~.
\ee 
\end{itemize}
The hypermultiplet composites (\ref{J-ABJM}), (\ref{L-ABJM}) and (\ref{A-ABJM}) are the components of the $\cN=6$ supercurrent. They are conserved 
as a consequence of the equations of motion (\ref{ABJM-eom}). It is interesting to note that these objects 
are obtained from the composites in \eqref{WW-eom} and 
\eqref{5.7} by taking the matrix trace.


\section{Prepotentials for the long $\cN=4$ vector multiplet}

For the left and right $\cN=4$ Yang-Mills supermultiplets, 
there exist {\it prepotential formulations}.
The harmonic superspace formulation was given by Zupnik  
\cite{Zupnik98,Zupnik99} in the case of $\cN=4$ Poincar\'e supersymmetry.
The projective superspace formulation was given in \cite{KT-M14} 
for the left  and right Yang-Mills  supermultiplets coupled to $\cN=4$ conformal supergravity (the case of abelian vector multiplets was described in \cite{KLT-M11}.
In this section we present a prepotential formulation for the long 
vector multiplet as a natural generalisation of Zupnik's construction 
\cite{Zupnik98,Zupnik99}.  A prepotential 
formulation for the long vector multiplet coupled to $\cN=4$ conformal supergravity 
may be obtained as a natural generalisation of the formulation 
developed in \cite{KT-M14}, but we will not elaborate on this here. 

Let $u^{\pm}_i$ and $v^{\pm}_{\tilde i}$ be
standard harmonic variables for the  $\sSU(2)_{\rm L}$ and $\sSU(2)_{\rm R}$,
\begin{subequations}
\bea
u^{+i} u^{-}_j - u^{-i} u^{+}_j &=& \delta^i_j~,\qquad\overline{u^{+i}} = u^-_i~,\\
v^{+\tilde i} v^{-}_{\tilde j} - v^{-\tilde i} v^{+}_{\tilde j}
  &=& \delta^{\tilde i}_{\tilde j}~,\qquad
\overline{v^{+\tilde i}} = v^-_{\tilde i}~.
\eea
\end{subequations}
The harmonics carry the labels $\pm$ which correspond to charges with respect to 
certain $\sU(1)$ subgroups of $\sSU(2)_{\rm L}$ and $\sSU(2)_{\rm R}$, 
respectively.  We will use these harmonics to parametrise  smooth superfields
on the harmonic superspace
\bea
 {\mathbb M}{}^{3|8} \times  \left[ {\sSU(2)}/{\sU(1)} \right]_{\rm L}
\times \left[ {\sSU(2)} / {\sU(1)} \right]_{\rm R} ~,
\label{biharmonic}
\eea
Any superfield $\F^{(p,q)}(z,u^\pm , v^\pm)$ defined on this superspace 
is labeled by two integer $\sU(1)$ charges $p $ and $q$ defined by 
$\F^{(p,q)}(z, \re^{\pm \ri \a} u^\pm , \re^{\pm \ri \b } v^\pm) 
= \re^{\ri (p\a + q \b )} \F^{(p,q)}(z,u^\pm , v^\pm)$, for real parameters $\a $ and $\b$.
It is useful to introduce left invariant vector fields for the groups
$\sSU(2)_{\rm L}$ and $\sSU(2)_{\rm R}$
\begin{subequations}
\label{harm-deriv}
\bea
&&D^{(2,0)} = u^+_i \frac\partial{\partial u^-_i}~,\quad
D^{(-2,0)} = u^-_i \frac\partial{\partial u^+_i}~,\quad
D^{(0,0)} = u^+_i \frac\partial{\partial u^+_i} - u^-_i \frac\partial{\partial u^-_i}~;
\\
&&D^{(0,2)} = v^+_{\tilde i} \frac\partial{\partial v^-_{\tilde i}}~,\quad
D^{(0,-2)} = v^-_{\tilde i} \frac\partial{\partial v^+_{\tilde i}}~,\quad
\tilde D^{(0,0)} = v^+_{\tilde i} \frac\partial{\partial v^+_{\tilde i}} - v^-_{\tilde i} \frac\partial{\partial v^-_{\tilde i}}~.
\eea
\end{subequations}
The operators within each of these sets obey the standard $\sSU(2)$ commutation relations 
\begin{subequations}
\bea
[D^{(0,0)}, D^{(\pm2,0)}] &=& \pm 2D^{(\pm2,0)}
~,\qquad
[D^{(2,0)},D^{(-2,0)}] = D^{(0,0)}~;
\label{38a}\\
{}
[\tilde D^{(0,0)}, D^{(0,\pm 2)}] &=& \pm 2D^{(0,\pm2)}~,\qquad
[D^{(0,2)},D^{(0,-2)}] = \tilde D^{(0,0)}~.
\label{38b}
\eea
\end{subequations}
Any two operators from the different sets commute with each other. 

We will work with matter multiplets  $\F^{(p,q)}(z,u^\pm , v^\pm)$ 
that transform under the gauge group as in eq. \eqref{3.1}. Since the gauge 
parameters $\t (z)$ in  \eqref{3.1} are harmonic independent, 
we now have a larger set of covariant derivatives
\bea
\cD_{\underline \cA} = (\cD_\cA , \cD^{(\pm 2, 0)}, \cD^{(0,0)}, \cD^{(0,\pm 2)}, 
\tilde{\cD}^{(0,0)} ) 
:= (\cD_\cA , D^{(\pm 2, 0)}, D^{(0,0)}, D^{(0,\pm 2)}, 
\tilde{D}^{(0,0)} ) ~~
\eea
possessing the gauge transformation 
\bea
{\cal D}'_{\underline \cA} &=& \re^{\ri \tau} {\cal D}_{\underline \cA} \re^{-\ri\tau }~.
\eea

We introduce a new basis for the
spinor gauge covariant derivatives\footnote{Switching off the gauge multiplet in \eqref{D-derivatives}
defines the new basis for the ordinary spinor covariant derivatives  ${D}^{i\tilde i}_\alpha$.}
${\cal D}^{i\tilde i}_\alpha$ 
and the gauge covariant field strengths ${\cal W}^{ij}$ and  ${\cal W}^{\tilde i \tilde j}$
as follows: 
\begin{subequations}
\bea
{\cal D}^{i\tilde i}_\alpha &\to &
{\cal D}^{(\pm1,\pm1)}_\alpha
 =u^\pm_i v^\pm_{\tilde i} {\cal D}^{i\tilde i}_\alpha~,\label{D-derivatives}\\
 {\cal W}^{ij}  &\to & ({\cal W}^{(2,0)},{\cal W}^{(-2,0)},{\cal W}^{(0,0)})
  = (u^+_i u^+_j, u^-_i u^-_j , u^+_i u^-_j){\cal W}^{ij}~, \label{6.7b}\\
{\cal W}^{\tilde i\tilde j}  & \to & ({\cal W}^{(0,2)},{\cal W}^{(0,-2)}, \tilde {\cal W}^{(0,0)})
  = (v^+_{\tilde i} v^+_{\tilde j}, v^-_{\tilde i} v^-_{\tilde j} , 
  v^+_{\tilde i} v^-_{\tilde j}){\cal W}^{\tilde i\tilde j}~. \label{6.7c}
\eea
\end{subequations}
Then the anti-commutation relation (\ref{N4-algebra}) leads to 
\begin{subequations} \label{D-algebra}
\bea
\{ {\cal D}^{(1,1)}_\alpha , {\cal D}^{(1,1)}_\beta \}
 &=&0~,
\label{6.8a}
\\
 \{ {\cal D}^{(1,1)}_\alpha , {\cal D}^{(-1,-1)}_\beta \} 
 &=&  2\ri {\cal D}_{\alpha\beta} -\varepsilon_{\alpha\beta}{\cal W}^{(0,0)}
 -\varepsilon_{\alpha\beta} \tilde{\cal W}^{(0,0)}~, \\
 \{ {\cal D}^{(1,-1)}_\alpha , {\cal D}^{(-1,1)}_\beta \} 
 &=& - 2\ri {\cal D}_{\alpha\beta} +\varepsilon_{\alpha\beta}{\cal W}^{(0,0)}
 -\varepsilon_{\alpha\beta} \tilde{\cal W}^{(0,0)}~,
 \eea
 \end{subequations}
 as well as to several additional relations which can be obtained from \eqref{D-algebra} by making use of the identities 
\begin{subequations}
 \bea
[\cD^{(2,0)},{\cal D}^{(1,\pm 1)}_\a ] &=&0~,\qquad
[ \cD^{(0,2)}, \cD^{(\pm 1,1) }_\a ] =0~,
\label{44} 
\\
\big[ \cD^{(-2,0)} , \cD_\a^{(1, \pm 1)} \big] &=& \cD_\a^{(-1,\pm 1)}~, \qquad 
  [ \cD^{(0,-2)} , \cD_\a^{(\pm 1,1)}] = \cD_\a^{(\pm 1, -1)}
 \eea
 \end{subequations}
 in conjunction with the relations
\begin{subequations} 
 \bea
 {\cal D}^{(2,0)}{\cal W}^{(0,0)}&=& \cW^{(2,0)}~, \qquad
{\cal D}^{(2,0)}{\cal W}^{(2,0)}=0~,\\
{\cal D}^{(0,2)} \tilde{\cW}^{(0,0)}&=& \cW^{(2,0)}~, 
\qquad
{\cal D}^{(0,2)}{\cal W}^{(0,2)}=0~.
\eea
\end{subequations}
 In particular, one obtains
\begin{subequations} \label{6.11}
\bea
 \{ {\cal D}^{(1,1)}_\alpha , {\cal D}^{(1,-1)}_\beta \}
 &=& -\varepsilon_{\alpha\beta}{\cal W}^{(2,0)}~,\quad
 \{ {\cal D}^{(1,1)}_\alpha , {\cal D}^{(-1,1)}_\beta \}
 = -\varepsilon_{\alpha\beta}{\cal W}^{(0,2)}~, 
 \\ 
 \{ {\cal D}^{(1,-1)}_\alpha , {\cal D}^{(-1,-1)}_\beta  \}
 &=& -\varepsilon_{\alpha\beta}{\cal W}^{(0,-2)}~,\quad
 \{ {\cal D}^{(-1,1)}_\alpha , {\cal D}^{(-1,-1)}_\beta  \}
 = -\varepsilon_{\alpha\beta}{\cal W}^{(-2,0)}~.~~~~~~
\eea
\end{subequations}
The Bianchi identities (\ref{W-constraints}) imply the analyticity constraints
 \bea
{\cal D}^{(1,\pm1)}_\alpha{\cal W}^{(2,0)}=0~,\quad
{\cal D}^{(\pm1,1)}_\alpha{\cal W}^{(0,2)}=0~.
\label{cov-analyt}
\eea

Remarkably, all information about the field strengths is encoded in the following equations:
\begin{subequations}  \label{6.13} 
\bea
{\cal D}^{(1,1)}_\alpha{\cal W}^{(2,0)}&=&0~, \qquad 
{\cal D}^{(2,0)}{\cal W}^{(2,0)}=0~, 
\qquad
{\cal D}^{(0,2)}{\cal W}^{(2,0)}=0~; \label{6.13a} \\
{\cal D}^{(1,1)}_\alpha{\cal W}^{(0,2)}&=&0~, \qquad 
{\cal D}^{(2,0)}{\cal W}^{(0,2)}=0~, 
\qquad
{\cal D}^{(0,2)}{\cal W}^{(0,2)}=0~.  \label{6.13b} 
\eea
\end{subequations}
Indeed, the third equation in \eqref{6.13a} tells us that ${\cal W}^{(2,0)}$
is independent of $v^\pm$, that is 
${\cal W}^{(2,0)}= {\cal W}^{(2,0)} (z, u^\pm)$. The second equation in \eqref{6.13a} tells us that ${\cal W}^{(2,0)} $ is independent of the harmonics $u^-_i$ 
and has the form \eqref{6.7b}. Finally, the first  equation in \eqref{6.13a} tells us that 
$\cW^{ij}$ obeys the Bianchi identity (\ref{W-constraints}). 

Eq.\ \eqref{6.8a} has two nontrivial implications. Firstly, it allows one to introduce 
 {\it covariantly semi-analytic} superfields  $\F^{(p,q)}(z,u^\pm , v^\pm)$  constrained by
\bea
\cD^{(1,1)}_\a \F^{(p,q)} = 0~.
\label{cs-a}
\eea
Such multiplets are rigid-superspace analogues of the covariant hybrid  multiplets
introduced in \cite{KLT-M11} in the framework of $\cN=4$ supergravity. 
Secondly, the constraint  \eqref{6.8a} has the following general solution 
\be
{\cal D}^{(1,1)}_\alpha 
 = \re^{-\ri\Omega} { D}^{(1,1)}_\alpha \re^{\ri\Omega}~,\qquad 
 \O \equiv \O^{(0,0)}~,
 \label{6.15}
 \ee
for some  bridge superfield $\O (z, u^\pm, v^\pm)$ which 
takes its values in the Lie algebra of the gauge group. 

Switching off the vector multiplet in \eqref{cs-a} defines {semi-analytic} superfields,
\be
D^{(1,1)}_\a \f^{(p,q)} = 0~.
\ee
In complete analogy with the harmonic superspace approach \cite{GIKOS,GIOS},
for such multiplets one can define a modified conjugation
 that maps every semi-analytic
 superfield $\f^{(p,q)}(z,u^\pm , v^\pm)$ into a semi-analytic one, 
 $\breve \f^{(p,q)}(z,u^\pm , v^\pm)$, of the same $\sU(1)$ charge.
 We will refer to it as ``smile-conjugation.''
The smile-conjugations has the property 
\be
\breve{\breve \f}^{(p,q)}(z,u^\pm , v^\pm)= (-1)^{p+q} \f^{(p,q)}(z,u^\pm , v^\pm)~. 
\ee
Thus in the case that $(p+q)$ is even, real semi-analytic superfields may be 
introduced. 

The introduction of $\O$ leads to a new  gauge freedom, 
in addition to the $\t$ gauge symmetry  \eqref{3.1}.
The gauge transformation of $\O$ is 
\begin{subequations}\label{6.1617}
\bea
\re^{\ri\Omega'} = \re^{\ri\lambda} \re^{\ri\Omega} \re^{-\ri\tau} ~,
\qquad \l \equiv \l^{(0,0)}~, 
\eea
where the new gauge parameter $\l (z, u^\pm, v^\pm)$ is constrained to be semi-analytic, 
\be
D^{(1,1)}_\alpha \lambda =0~,
\label{hybrid-constraint}
\ee
\end{subequations}
and is real with respect to the smile-conjugation. The bridge $\O$ in \eqref{6.15}
may be chosen Hermitian with respect to the operations of transposition and
smile-conjugation. 

Making use of the bridge allows us to introduce a new representation 
for the covariant derivatives $ \cD_{\underline \cA} $
and matter multiplets $\F^{(p,q)}$ with the property that no $\t$-gauge freedom
is left. It is obtained by applying the transformation:
\begin{subequations} \label{5.199}
 \bea
 \cD_{\underline \cA} ~ &\to&  ~ 
 \nabla_{\underline \cA} =
\re^{\ri\Omega} \cD_{\underline \cA}  \re^{-\ri\Omega} ~,\\
\F^{(p,q)}  ~&\to & ~\f^{(p,q)} = \re^{\ri\Omega}\F^{(p,q)} ~,
\eea
in particular
\bea
{\cal W}^{(p,q)} ~&\to & ~W^{(p,q)}= \re^{\ri\Omega}{\cal W}^{(p,q)} \re^{-\ri\Omega}~.
\eea
\end{subequations}
for the field strengths \eqref{6.7b} and \eqref{6.7c}. 
The resulting $\l$-representation is characterised by two important properties.
Firstly, 
the covariant derivative $\nabla^{(1,1)}_\alpha$ has no gauge connection,
\be
\nabla^{(1,1)}_\alpha = D^{(1,1)}_\alpha~.
\ee
Secondly, 
 the harmonic covariant derivatives 
 acquire gauge connections,
\be
\nabla^{(\pm2,0)} = D^{(\pm2,0)}+\ri V^{(\pm2,0)}~,\qquad
\nabla^{(0,\pm2)} = D^{(0,\pm2)}+\ri V^{(0,\pm2)}~.
\label{gauge-cov-harm}
\ee
Under the $\lambda$-gauge transformation,
the gauge connections in (\ref{gauge-cov-harm}) change as
\be
\ri V'^{(\pm2,0)} =  \re^{\ri\lambda }( \nabla^{(\pm2,0)} e^{-\ri\lambda})~,\qquad
\ri V'^{(0,\pm2)} =  \re^{\ri\lambda }( \nabla^{(0,\pm2)} \re^{-\ri\lambda})~.
\ee
In the $\lambda$-representation, 
the equations (\ref{44}) mean that the gauge prepotentials $V^{(2,0)}$ and $V^{(0,2)}$ are semi-analytic, 
\be
D^{(1,1)}_\a V^{(2,0)} =0 ~,\qquad
D^{(1,1)}_\a V^{(0,2)} =0 ~.
\label{hybrid-constr}
\ee

The above consideration in this section concerns
the long vector multiplet. As discussed in section \ref{SectN4}, 
the left and the right vector multiplets are obtained from the long one  
by imposing the additional constraints \eqref{3.4a}
and \eqref{3.4b}, respectively. This  leads to important specific features, 
which we now analyse. 
It suffices to consider only the left multiplet for which 
${\cal W}^{\tilde i \tilde j} =0$, and hence 
${\cal W}^{(0,2)} = {\cal W}^{(0,-2)} =  \tilde {\cal W}^{(0,0)} =0$; 
the case of the right vector multiplet is analogous.  
Since the right-hand sides of eqs. \eqref{D-algebra} and \eqref{6.11}
are independent of the $v^\pm$ harmonics, the bridge $\O$ in \eqref{6.15}
can also be chosen to be independent of these harmonics, 
$\O = \O (z, u^\pm)$. The gauge parameter $\l$ in \eqref{6.1617}
also becomes  independent of the $v^\pm$ harmonics,
$\l = \l (z, u^\pm)$, and the semi-analyticity constraint \eqref{hybrid-constraint}
turns into the analyticity conditions $D^{(1,\pm1)}_\alpha \l=0$. 
These results have two important corollaries:
(i) the harmonic connections  $V^{(0,\pm2)}$ in \eqref{gauge-cov-harm}
vanish, $V^{(0,\pm2)} =0$; (ii) the connections $V^{(\pm2,0)}$ are independent 
of the $v^\pm $ harmonics, $V^{(\pm2,0)} = V^{(\pm2,0)} (z, u^\pm)$. 
As a result, the first equation in \eqref{hybrid-constr}
obeys the stronger analyticity conditions $D^{(1,\pm1)}_\alpha V^{(2,0)}=0$, 
which agrees with Zupnik's approach \cite{Zupnik98,Zupnik99}.

The zero-curvature conditions (\ref{38a}) and (\ref{38b}) in the $\lambda$-frame are
\begin{subequations}
\bea
D^{(2,0)}V^{(-2,0)} - D^{(-2,0)} V^{(2,0)} +\ri[V^{(2,0)},V^{(-2,0)}] &=&0~,
\\
D^{(0,2)}V^{(0,-2)} - D^{(0,-2)} V^{(0,2)} +\ri[V^{(0,2)},V^{(0,-2)}]&=&0~.
\eea
\end{subequations}
They allow one to express the superfields $V^{(-2,0)}$ and $V^{(0,-2)}$ in terms of 
$V^{(2,0)}$ and $V^{(0,2)}$, respectively,
\be
V^{(-2,0)} = V^{(-2,0)}\big[ V^{(2,0)}\big]~,\qquad
V^{(0,-2)} = V^{(0,-2)}\big[ V^{(0,2)}\big]~.
\ee
Explicitly, these solutions are given as series over harmonic distributions presented in \cite{Zupnik87}. We point out that the superfields $V^{(-2,0)}$ and $V^{(0,-2)}$ live in the full superspace in contrast to the  prepotentials $V^{(2,0)}$ and $V^{(0,2)}$ subject to the constraints 
(\ref{hybrid-constr}).

In the case of the left vector multiplet, for which ${\cal W}^{\tilde i \tilde j} =0$ and 
$V^{(0,\pm2)} =0$, the prepotential $V^{(2,0)}$ is real analytic but otherwise 
unconstrained. For the long vector multiplet, we have two semi-analytic prepotentials 
 $V^{(2,0)}$ and $V^{(0,2)}$. The constraints (\ref{hybrid-constr}) are not the only 
 conditions they obey. They are also related to each other by the zero-curvature condition
\bea
D^{(2,0)}V^{(0,2)} - D^{(0,2)} V^{(2,0)} +\ri[V^{(2,0)},V^{(0,2)}] &=&0~.
\label{6.26} 
\eea

Using the algebra of spinor covariant derivatives in the $\lambda$-frame it is possible to express the field strengths $ W^{(2,0)}$ and $ W^{(0,2)}$ in terms of the gauge prepotentials $V^{(0,-2)}$ and $V^{(-2,0)}$. The resulting expressions are
\be
W^{(2,0)} = \frac\ri2 D^{(1,1)\alpha} D^{(1,1)}_\alpha V^{(0,-2)}~,
\qquad
W^{(0,2)} = \frac\ri2 D^{(1,1)\alpha} D^{(1,1)}_\alpha V^{(-2,0)}~.
\label{WV}
\ee
These superfields transform covariantly under the $\lambda$-gauge group, 
\be
W'^{(2,0)} = \re^{\ri\lambda}W^{(2,0)} \re^{-\ri\lambda}~,\qquad
W'^{(0,2)} = \re^{\ri\lambda}W^{(0,2)} \re^{-\ri\lambda}~.
\ee
These results show that the long vector multiplet is completely described in terms of 
the two real semi-analytic prepotentials  $V^{(2,0)}$ and $V^{(0,2)}$
subject to the zero-curvature condition \eqref{6.26}.

As discussed above, all information about the off-shell long vector 
multiplet is encoded in the equations \eqref{6.13}. 
It is interesting that the on-shell hypermultiplets are described by analogous equations:
\begin{subequations}  \label{6.hyp} 
\bea
{\cal D}^{(1,1)}_\alpha {q}^{(1,0)}&=&0~, \qquad 
{\cal D}^{(2,0)}{q}^{(1,0)}=0~, 
\qquad
{\cal D}^{(0,2)}{q}^{(1,0)}=0~; \label{6.hypa} \\
{\cal D}^{(1,1)}_\alpha {q}^{(0,1)}&=&0~, \qquad 
{\cal D}^{(2,0)}{q}^{(0,1)}=0~, 
\qquad
{\cal D}^{(0,2)}{q}^{(0,1)}=0~.  \label{6.hypb} 
\eea
\end{subequations}
For example, consider the left hypermultiplet. 
The third equation in \eqref{6.hypa} tells us that ${q}^{(1,0)}$
is independent of $v^\pm$, that is ${q}^{(1,0)}= {q}^{(1,0)} (z, u^\pm)$. 
The second equation in \eqref{6.hypa} tells us that ${q}^{(1,0)} $ is independent of the harmonics $u^-_i$ 
and has the functional form $q^{(1,0)} = u^{+}_i q^i$. 
Finally, the first  equation in \eqref{6.hypa} tells us that 
$q^{i}$ obeys the constraint \eqref{9a}. 
In summary, the on-shell hypermultiplets in harmonic superspace are described by the superfields
\be
q^{(1,0)} = u^{+}_i q^i~,\qquad
q^{(0,1)} = v^{+}_{\tilde i} q^{\tilde i}~.
\ee

In conclusion of this section let us briefly discuss the equations of motion 
for  the ABJM theory in $\cN=4$ harmonic superspace. 
This theory is described by the hypermultiplet superfields $q^{(0,1)}$ and $q^{(1,0)}$ in the bi-fundamental representation of the gauge group $G_{\frak L}\times G_{\frak R}$. There are also superfield strengths $ W^{(2,0)}_{\frak L}$, $ W^{(2,0)}_{\frak R}$, $ W^{(0,2)}_{\frak L}$ and $ W^{(0,2)}_{\frak R}$ which take values in the Lie algebras of the gauge groups $G_{\frak L}$ and $G_{\frak R}$. 
These superfields obey the equations  \eqref{6.13} and \eqref{6.hyp}.
The field strengths are expressed in terms of the hypermultiplets as 
\bea
&&W_{\frak L}^{(2,0)} = \ri \kappa\, q^{(1,0)}\bar q^{(1,0)}~,\quad
W_{\frak R}^{(2,0)} = \ri \kappa\, \bar q^{(1,0)} q^{(1,0)}~,\non\\
&&W_{\frak L}^{(0,2)} = \ri \kappa\, q^{(0,1)}\bar q^{(0,1)}~,\quad
W_{\frak R}^{(0,2)} = \ri \kappa\, \bar q^{(0,1)} q^{(0,1)}~.
\eea
It would be interesting to find a superfield Lagrangian reproducing this set of equations.


\section{Conclusion} 

We have shown that the  $\cN=4$ superfield realisations for 
superconformal CS theories with $\cN\geq 4$
require the long $\cN=4$ vector multiplet.
The structure of the long $\cN=4$ vector multiplet turns out to be 
the main reason for problems with constructing off-shell actions 
in $\cN=4$ superspace 
for supersymmetric CS theories with eight and more supercharges. 
The simplest way to see this is to look at $\cN=4 \to \cN=3$ superspace 
reduction of large $\cN=4$ vector multiplet. In $\cN=3$ superspace, 
this multiplet is described by gauge covariant symmetric isospinors 
 ${\mathbb W}^{ij}$ and $\tilde{\mathbb W}^{ij}$  in the adjoint representation of the gauge group.
One of them, ${\mathbb W}^{ij}$, is the field strength of  
the $\cN=3$ vector multiplet \cite{ZH,Zupnik98}.  
In terms of the gauge covariant derivatives ${\cal D}^{ij}_\alpha$
in $\cN=3 $ superspace, it originates as follows
\be
\{ {\cal D}^{ij}_\alpha, {\cal D}_\beta^{kl} \}
= -2\ri\varepsilon^{i(k}
\varepsilon^{l)j} {\cal D}_{\alpha\beta} 
+\frac12\varepsilon_{\alpha\beta}
\big(\varepsilon^{i(k}{\mathbb W}^{l)j} + \varepsilon^{j(k}{\mathbb W}^{l)i} \big)
\label{W-algebra}
\ee
and obeys the Bianchi identity
\be
{\cal D}^{(ij}_\alpha{\mathbb W}^{kl)} =0~.
\ee
The other object, $\tilde{\mathbb W}^{ij}$, is a Lie-algebra-valued 
matter multiplet subject to the same constraint as ${\mathbb W}^{ij}$, 
\be
{\cal D}^{(ij}_\alpha \tilde{\mathbb W}^{kl)} =0~.
\ee
Each of  ${\mathbb W}^{ij}$ and $\tilde{\mathbb W}^{ij}$ 
is a  linear combination of the $\cN=4$ field strengths  $\cW^{ij}$ and 
$ \cW^{\tilde i \tilde j} $ in \eqref{N4-algebra} 
projected to $\cN=3$ superspace. On the mass shell, 
${\mathbb W}^{ij}$ and $\tilde{\mathbb W}^{ij}$ become composites
constructed from $\cN=3$ hypermultiplets $q^i$ and their conjugates $\bar q_i$.
Symbolically, we have ${\mathbb W}_A^{ij} = \bar q^{(i}  T_A q^{j)} $ 
and $\tilde{\mathbb W}_A^{ij}=  \bar q^{(i}   \tilde{T}_A q^{j)} $.
The former equation can always be realised as the equation of motion for the vector 
multiplet in some $\cN=3$ superconformal Chern-Simons-matter theory 
formulated in harmonic superspace \cite{BILPSZ}.
However, there is no systematic way to realise the latter constraint as an 
Euler-Lagrange equation, except for the abelian case. 

The results of this paper can naturally be generalised to supergravity.
The equations of motion for a general $\cN=4$ superconformal CS theory 
coupled to $\cN=4$ conformal supergravity are
\begin{subequations}\label{6.33}
\bea
{\bm \nabla}_\alpha^{\tilde i (i}q^{j)} &=& 0~,  
\qquad   
{\bm \nabla}_\alpha^{i(\tilde i }q^{\tilde j)} = 0 ~, 
\label{6.33a}  \\
\frac{1}{\k} \cW^{ij}_A &=& \ri \,g_{AB}\, \bar q^{(i} T^B  q^{j)}~,
\qquad
\frac{1}{\tilde \k}\cW^{\tilde i \tilde j}_A = \ri 
\,g_{AB}\, \bar q^{(\tilde i} \tilde T^B 
q^{\tilde j)} ~, \\ 
\frac{1}{\k_{\rm SG}}W &=&   \bar q_i  q^i -  \bar q_{\tilde i} q{}^{\tilde i} ~, 
\label{csgem}
\eea
\end{subequations}
where $W$ is the $\cN=4$ super-Cotton scalar (see \cite{BKNT-M1} for more details).
The torsion and curvature tensors in $\cN=4$ conformal superspace are completely determined in terms of $W$ and its covariant derivatives. The super-Cotton scalar
obeys the Bianchi identity \cite{BKNT-M1} 
\bea
\nabla^{\alpha\, (\tilde i (i } \nabla_\alpha^{j)\tilde j )} W =0~,  
\eea
and the same equation is obeyed by each term on the right of \eqref{csgem}.
The gauge-covariant derivative  ${\bm\nabla}$ in eq. \eqref{6.33a}
is defined in Appendix B, eq. \eqref{B.1}. In this paragraph as well as in Appendix B, 
we use the notation $\nabla_\cA $ for the covariant derivatives in conformal superspace. These  should not be confused with the gauge covariant 
derivative in the $\l$-frame \eqref{5.199}.
\\

\noindent
{\bf Acknowledgements:}\\
We are grateful to Joseph Novak for reading the manuscript and verifying the consistency of \eqref{csgem}.
This work is supported in part by the Australian Research Council, project No.
 DP140103925.


\appendix

\section{Consistency condition in $\cN=4$ CS theories}
\label{AppA}

As shown in section \ref{SectN4}, 
the $\cN=4$ supersymmetric gauge theories are 
subject to the consistency conditions (\ref{int-cond}). 
In the case of superconformal CS theories,
these conditions imply the equations (\ref{15}) for the generators of the gauge group. Here we demonstrate that these equations are equivalent to the fundamental identity for the generators of the gauge group found in \cite{GW}. 
For simplicity, here we consider only the left hypermultiplet $q^i$ and field strength $\cW^{ij}$;
the analysis for the right multiplets 
$q^{\tilde i}$ and $\cW^{\tilde i \tilde j}$
is absolutely identical.

We can view the hypermultiplet $(q^i)=q^i_a$, $a=1,\ldots,n$, as a $n$-vector in some representation of the gauge group $G$ so that $(\bar q^i)=\bar q^{i a}$ form the conjugated representation.
It is convenient to combine them into one $2n$-dimensional vector 
$Q^i= (Q^i_{\hat a} )= (q^i_a, - \bar q^{ia})^{\rm T}$, where  $\hat a=1,\ldots,2n$, 
corresponds to the $\sSp(2n)$ group such that
\be
 \bar Q_i^{\hat a} :=
\overline{Q^i_{\hat a}} =\varepsilon_{ij} \Omega^{\hat a\hat b} Q^j_{\hat b}~, 
\qquad  \bar Q_i = ( \bar Q_i^{\hat a} )=(\bar q_i^a , q_{ia})~,
\label{Q}
\ee
where $\Omega^{\hat a\hat b}=-\Omega^{\hat b\hat a}$ is the invariant tensor of $\sSp(2n)$, 
\be
\Omega_{\hat a\hat b} = \left(
\begin{array}{cc}
0 & {\mathbbm 1}_{n}\\
-{\mathbbm 1}_n &0
\end{array}
\right).
\ee 
The gauge group $G$ acts on $Q^i$ by symplectic transformations. Denoting by  
$({\bf T}^A)_{\hat a}{}^{\hat b}$ the generators of $G$ 
in which $Q^i_{\hat a}$ transforms, the field strength is
\be
{\cal W}^{ij} ={\cal W}^{ij}_A {\bf T}^A~.
\ee
In this notation, the integrability condition (\ref{int-cond}) can be rewritten as
\be
{\cal W}_A^{(ij} ({\bf T}^A)_{\hat a}{}^{\hat b} Q^{k)}_{\hat b}=0~.
\label{constraint2}
\ee

In the superconformal CS theories considered in section \ref{SectN4},
the field strength ${\cal W}^{ij}_A$
becomes the hypermultiplet composite operator given by (\ref{W}). 
In the notation (\ref{Q}), eq. (\ref{W}) reads
\be
{\cal W}^{ij}_A = \frac\ri2 \kappa\, g_{AB}\, Q^{\hat a(i} ({\bf T}^B)_{\hat a}{}^{\hat b}  Q^{j)}_{\hat b}~,
\ee
where the generator $({\bf T}^A)_{\hat a}{}^{\hat b}$ has the following block diagonal form
\be
({\bf T}^A)_{\hat a}{}^{\hat b} = 
\left(
\begin{array}{cc}
(T^A)_a{}^b & 0 \\
0 & - (T^A)_b{}^a
\end{array}
\right).
\ee

The consistency condition 
\eqref{constraint2} now reads
\be
g_{AB} Q^{\hat a(i} Q^j_{\hat b} Q^{k)}_{\hat d} ({\bf T}^A)_{\hat a}{}^{\hat b} ({\bf T}^B)_{\hat c}{}^{\hat d} =0~.
\ee
This equation implies the following constraint on the generators of the gauge group
\be
g_{AB} {\bf T}^A_{\hat a(\hat b} {\bf T}^B_{\hat c\hat d)} =0~,
\label{fund-identity}
\ee
where we have assumed that the hypermultiplet indices, $\hat a$, $\hat b , \dots$,   
are raised and lowered 
using the symplectic metric $\Omega^{\hat a\hat b} $ and its inverse $\Omega_{\hat a\hat b} $,
$\Omega_{\hat a\hat b}\Omega^{\hat b\hat c} = \delta_{\hat a}^{\hat c}$.
For the generators with lower indices, 
${\bf T}^A_{\hat a\hat b} = \Omega_{\hat b\hat c} ({\bf T}^A)_{\hat a}{}^{\hat c}$, 
we have 
\be
({\bf T}^A)_{\hat a\hat b} = 
\left(
\begin{array}{cc}
0 & (T^A)_a{}^b  \\
 (T^A)_b{}^a &0
\end{array}
\right).
\ee
With this block matrix representation of the generators it becomes obvious that 
eq. (\ref{fund-identity}) is equivalent to (\ref{15a}),
\be
g_{AB} {\bf T}^A_{\hat a(\hat b} {\bf T}^B_{\hat c\hat d)} =0
\quad\Longleftrightarrow\quad
g_{AB}(T^A)_a{}^{(b} (T^B)_c{}^{d)} =0~.
\ee

The equation (\ref{fund-identity}) was first derived in \cite{GW} 
where the formulation in terms of  $\cN=1$ superfields was developed for 
$\cN=4$ superconformal CS theories. Eq.\ (\ref{fund-identity})
was necessary for the construction of consistent interaction Lagrangians for $\cN=4$ superconformal CS theories. As demonstrated in our paper,
in  $\cN=4$ superspace eq.\ (\ref{fund-identity}) naturally arises as the consistency condition of the hypermultiplet equations of motion.

In \cite{GW}, the equation (\ref{fund-identity}) was named as the {\it fundamental identity} since it imposes non-trivial  constraints both on the gauge group and the matter representation. 
In the same paper it was demonstrated that the fundamental identity is satisfied for those Lie groups which allow for super-extensions. The typical examples of such gauge groups are $\sU(M)\times \sU(N)$ and $\sO(M)\times \sSp(2N) $ which are the bosonic bodies of $\sU(M|N)$ and $\sOSp(M|2N)$, respectively. The matter hypermultiplets belong to the bi-fundamental representations of these gauge groups.


\section{$\cN$-extended vector multiplet}\label{AppendB}

To describe a Yang-Mills multiplet in the 3D 
$\cN$-extended conformal superspace $\cM^{3|2\cN}$ of \cite{BKNT-M1}, 
 parametrized by coordinates $z^M = (x^m, \ \q^\mu_I)$, 
we introduce gauge covariant derivatives
\bea
 {\bm\nabla}_\cA =( {\bm\nabla}_a ,  {\bm\nabla}_\a^I)
 := \nabla_\cA + \ri V_\cA~, \qquad I=1, \dots, \cN~, 
 \label{B.1}
\eea
where $\nabla_\cA$ are the supergravity covariant derivatives \cite{BKNT-M1}.
The algebra of gauge covariant derivative is
\begin{align} 
[{\bm \nabla}_\cA, {\bm \nabla}_\cB\} &= 
 \ri \cF_{\cA \cB} + \dots \ ,
\end{align}
where the ellipsis denotes the purely supergravity terms. 
The field strength $\cF_{\cA \cB}$ 
satisfies the Bianchi identity
\be 
 \bm \nabla_{[\cA} \cF_{\cB \cC\}} + T_{[\cA \cB}{}^\cD \cF_{|\cD| \cC\}} = 0~, 
\ee
where $T_{\cA \cB}{}^\cD$ is the torsion tensor, see \cite{BKNT-M1} 
for more details. The field strength 
is subject to  a covariant constraint to describe a vector multiplet. 
For  $\cN>1$ the constraint \cite{HitchinKLR,ZP,ZH} is 
\be
\cF_{\a}^I{}_\b^J =  2\ri\ve_{\a\b}\cW^{IJ} \ , \label{Fconst}
\ee
The Bianchi identities then give the remaining components
of the field strength  \cite{BKNT-M1}:
\bsubeq \label{FSComps}
\bea
\cF_{a}{}_\a^I&=&
-\frac{1}{ (\cN-1)}(\g_a)_\a{}^{\b} \bm \nabla_{\b J} \cW^{I J}
~,
\\
\cF_{ab}&=&
\frac{\ri}{ 4\cN(\cN-1)}\ve_{abc}(\g^c)^{\a\b}[\bm \nabla_{\a}^{ K}, \bm \nabla_{\b}^{ L}] \cW_{ K L}
~.
\eea
\esubeq
For  $\cN>2$ the field strength $\cW^{I J}$ 
is constrained by the dimension-3/2 Bianchi identity 
\bea
\bm \nabla_{\g}^{I} \cW^{ J K}&=&
\bm \nabla_{\g}^{[I} \cW^{ J K]}
- \frac{2}{ \cN-1} \d^{I [J} \bm \nabla_{\g L} \cW^{ K] L}
~.
\label{VMBI}
\eea
This constraint may be shown to define an off-shell 
supermultiplet \cite{GGHN}, see also \cite{BKNT-M1}.

The component fields of vector multiplets may be extracted from the 
field strength $\cW^{IJ}$. 
For $\cN > 1$, we define the matter fields as follows
\allowdisplaybreaks{
\bsubeq \label{components}
\begin{align}
w^{IJ} &:= \cW^{IJ}| \ , \\
\l_\a^I &:= \frac{2}{\cN - 1} \bm \nabla_{\a J} \cW^{IJ}| \ , \\
h^{IJ} &:= \frac{\ri}{\cN - 1} \bm \nabla^{\g [I} \bm \nabla_{\g K} \cW^{J] K}| \ , \\
\c_{\a_1 \cdots \a_n}{}^{I_1 \cdots I_{n+2}} &:= I(n) \bm \nabla_{(\a_1}^{[I_1} \cdots \bm \nabla_{\a_n)}^{I_n} \cW^{I_{n+1} I_{n+2}]}| \ ,
\end{align}
\esubeq}
where
\be \label{Ifunct}
I(n)=
\begin{cases}
\ri \ , & n = 1,2 \ ({\rm mod} \ 4) \\
1 \ , & n = 3,4 \ ({\rm mod} \ 4) ~.
\end{cases}
\ee
For the supersymmetry transformations of the $\cN$-extended vector multiplet, see 
\cite{KN14}.

The component fields of the vector multiplet 
form the following tower \cite{GGHN,KN14}:\\
\begin{minipage}[t]{\textwidth}
\begin{picture}(430,195)
\put(197,170){$w^{I J}$}
\put(205,165){\vector(-1,-1){20}}
\put(210,165){\vector(1,-1){20}}
\put(165,130){$\c_\a{}^{I J K}$}
\put(230,130){$\l_\a{}^{I}$}
\put(165,125){\vector(-1,-1){20}}
\put(175,125){\vector(1,-1){20}}
\put(240,125){\vector(-1,-1){20}}
\put(250,125){\vector(1,-1){20}}
\put(125,90){$\c_{\a_1\a_2}{}^{I_1 \cdots I_4}$}
\put(197,90){$h^{I J}$}
\put(270,90){${F}_{\a_1 \a_2}$}
\put(122,80){\vector(-1,-1){20}}
\put(83,45){$\cdots$}
\put(80,38){\vector(-1,-1){20}}
\put(40,0){$\c_{\a_1\cdots\a_{\cN-2}}{}^{I_1\cdots I_{\cN}}$
}
\end{picture}
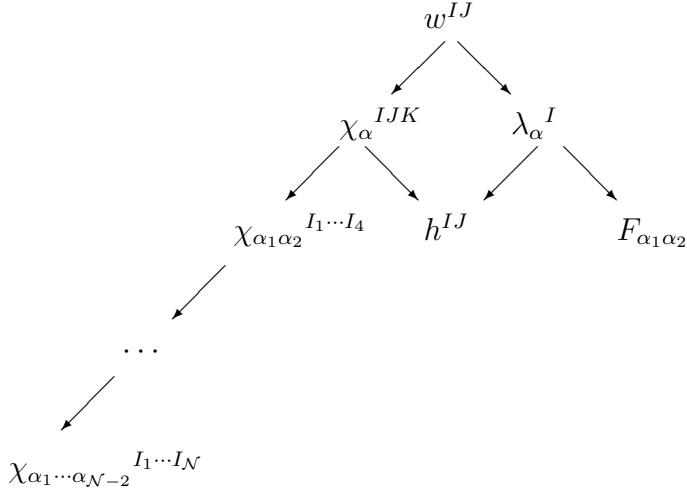
\begin{center}{\bf Figure 1.} Component fields of the $\cN$-extended vector multiplet\end{center}
\end{minipage} \\

\noindent
Here $F_{\a\b}$ is the symmetric spinor associated 
with the Hodge dual  $F^a = \frac{1}{2} \ve^{abc} F_{bc}$ of
 $F_{ab} = \cF_{ab}|$.
Modulo fermionic terms, $F_{ab}$ coincides with the component field strength.
In the left branch of the diagram, the fields $\c_{\a_1\a_2}{}^{I_1 \cdots I_4}, \dots,
\c_{\a_1\cdots\a_{\cN-2}}{}^{I_1\cdots I_{\cN}}
$ satisfy, in the linearised approximation,
the conservation equations 
$(\g^a)^{\b\g}\pa_a\c_{\b\g\a_3\cdots\a_n}{}^{I_1\cdots I_{n+2}}=0$.

In the $\cN=4$ case, the two branches of the diagram have identical algebraic structure, 
and thus every field on the left has a twin on the right. This doubling of fields 
disappears if the field strength $\cW^{IJ}$ is constrained to be self-dual,
$\tilde{\cW}^{IJ} = {\cW}^{IJ} $, or anti-self-dual, $\tilde{\cW}^{IJ} = -{\cW}^{IJ}$,  
where
\be
\tilde{\cW}^{IJ} := \hf \ve^{IJKL}\cW_{KL} \ .
\ee
These cases correspond to the left and right vector multiplets, respectively.

\begin{footnotesize}

\end{footnotesize}

\end{document}